\documentstyle[12pt]{article}
\begin{document}

\vskip 1truecm

\begin{center}
{\LARGE \bf Construction of Sources for\\ 
\vspace{0.5cm}Majumdar-Papapetrou Spacetimes}

\vskip .4truecm

{\bf Victor Varela}

Escuela de F\'{\i}sica, Facultad de Ciencias,\\
Universidad Central de Venezuela,\\
Caracas, Venezuela \\
{\em E-mail: {\tt vvarela@fisica.ciens.ucv.ve}}\\

\end{center}

\abstract{
We study Majumdar-Papapetrou solutions
for the 3+1 Einstein-Maxwell equations, with charged dust acting as
the external source for the fields. The spherically symmetric solution
of G\"{u}rses is considered in detail. We introduce new parameters that
simplify the construction of class $C^1$,  singularity-free geometries.
The arising sources are bounded or unbounded, and the redshift of light
signals allows an observer at spatial infinity to distinguish these cases.
We find out an interesting affinity between the conformastatic metric
and some homothetic, matter and Ricci collineations. The associated
non-Noetherian symmetries provide us with distinctive solutions that can 
be used
to construct non-singular sources for Majumdar-Papapetrou spacetimes.}

\vskip 1truecm

\section{Introduction}
Majumdar \cite{ma} and Papapetrou \cite{pa} showed that assuming a static
spacetime metric
with the conformastatic \cite{syn} form
\begin{equation}
ds^{2}=-V^2\;dt^{2}+\frac{1}{V^2}\;d\vec{x}\cdot d\vec{x},
\label{metric}
\end{equation}
where $V=V\left(x^{1},x^{2},x^{3}\right)$,
the task of finding solutions for the Einstein-Maxwell equations can be
greatly simplified. In fact, assuming a linear relationship between
the time-like component of the electromagnetic potential and
$V$, they found that $\frac{1}{V}$ is harmonic, i.e. it is a
solution of the Laplace equation. Therefore, the problem of solving the
coupled field equations reduces to solving the Laplace equation in a
three-dimensional, Euclidean space which is called the "background space".
Solutions with point-like singularities distributed in a bounded region
of the background
space imply asymptotically flat spacetimes and have the simple
form

\begin{equation}
\frac{1}{V}=1+\sum\limits_{i=1}^{N}\frac{m_{i}}{r_{i}},
\label{mpsol}
\end{equation}
where
\begin{equation}
r_{i}=\sqrt{(x-x_{i})^{2}+(y-y_{i})^{2}+(z-z_{i})^{2}}.
\label{euclidist}
\end{equation}

A complete analysis of the singularities of these electrovac
solutions was presented by
Hartle and Hawking \cite{haha} who showed that in a static system of 
electric
charges of the same sign, all the genuine spacetime singularities are
contained within event horizons. This class of electrovac solutions
describe systems of charged black holes in equlibrium under their mutual
gravitational and electrical interactions.

The singularities of the Majumdar-Papapetrou (MP) solution (\ref{mpsol}) can
be avoided in principle if we consider extended sources for the electric and
gravitational fields e.g. charged fluids. Das \cite{das} discussed
the extension
of the MP procedure to the case in which the fluid is charged dust.
As a result, $\frac{1}{V}$ comes out as
the solution of a potential equation which is generally non-linear
and additional information about the energy density of the fluid, $\rho$,
must be provided in order to obtain particular solutions.

G\"{u}rses \cite{gur} showed how a particular choice
for the functional $\rho=\rho\left( V \right)$ makes the potential 
equation linear and
leads to simple analytical, internal solutions that
can be matched with the external
solution (\ref{mpsol}) at the boundary of the charged source, without 
symmetry restrictions.
Other choices for $\rho$
were considered by Varela \cite{var}, who discussed charged dust solutions
in the non-linear case of this equation. Ida \cite{ida} studied solutions
with charged perfect fluids, and Cho {\it et al.} \cite{cho} generalised 
this theory
with the inclusion of a dilaton field.

A different approach was presented by
Bonnor and Wickramasuriya \cite{bw1,bw2} and Bonnor \cite{bo1}, who 
constructed non-singular
charged dust sources with spherical and spheroidal shapes. They used a 
good deal of analytical
intuition to obtain the corresponding metrics without solving the 
non-linear potential
equation. These and other electro-gravitational solutions
are reviewed in the recent work by Ivanov \cite{iva}.

The G\"{u}rses method uses the power of linearity for the construction
of sources for MP spacetimes. However, new singularities can be included 
in the geometry
when the parameters take arbitrary values.
This fact is simply illustrated by his spherically symmetric solution,
which is prone to develope periodic singularities.

This paper is organised as follows: in the next Section we consider 
fundamental facts
of the MP approach and review the proof that the relationship between 
the electrostatic potential
and $V$ is actually a consequence of the field equations. This is 
followed by a discussion
of the electrovac MP solution in the spherically symmetric case. Section 
III looks at the
G\"{u}rses asymptotically flat, class $C^1$  geometry. We define new 
parameters that allow
a careful choice of families of non-singular solutions. Unbounded dust 
sources are examined in
Section IV, where solutions of the non-linear potential equation are 
matched to the G\"{u}rses
internal geometry. In Section V we investigate MP sources with 
non-Noetherian symmetries.
These symmetries are formulated in terms of Lie derivatives of
the metric and energy-momentum tensors, and provide us with solutions 
obtained in a very
elegant manner. This approach dispenses with the need for {\it ad-hoc} 
fuctionals
$\rho=\rho(V)$ and confronts us with distinctive classes of MP 
solutions. We conclude in
Section VI with a brief discussion of our results.

\section{Majumdar-Papapetrou Solutions}
We consider solutions for the Einstein-Maxwell-charged dust (EMCD) equations
\begin{equation}
G_{\mu\nu}=8\pi T_{\mu\nu},
\label{eins}
\end{equation}
\begin{equation}
F^{\mu\nu}{}_{;\nu}=4\pi J^{\mu},
\label{max}
\end{equation}
where $G_{\mu\nu}$ and $F^{\mu\nu}$ denote
the Einstein and Maxwell tensors,  and the total
energy-momentum tensor is given by
\begin{equation}
T_{\mu\nu}=E_{\mu\nu}+M_{\mu\nu}.
\label{temunu}
\end{equation}
The Maxwell energy-momentum tensor is
\begin{equation}
E_{\mu\nu}=\frac{1}{4\pi}\left(F_{\mu\alpha}F_{\nu}^{ \alpha}-
\frac{1}{4}g_{\mu\nu}F_{\alpha\beta}F^{\alpha\beta}\right),
\label{maxtmunu}
\end{equation}
and the matter term
\begin{equation}
M_{\mu\nu}=\rho u_{\mu}u_{\nu}
\label{dusttmunu}
\end{equation}
corresponds to
dust with energy density $\rho$ and four-velocity $u^{\mu}$.
The four-current is defined by the expression
\begin{equation}
J^{\mu}=\sigma u^{\mu},
\label{current}
\end{equation}
where $\sigma$ is the charge density.

We assume a static spacetime and use the conformastatic metric
\begin{equation}
ds^{2}=-V^2\;dt^{2}+\frac{1}{V^2}\;h_{ij}dx^{i}dx^{j},
\label{confor}
\end{equation}
where the background metric $h_{ij}$ and $V$ depend only on the space-like
coordinates $x^{1}$, $x^{2}$, $x^{3}$.
The electrostatic forms of $A_{\mu}$ and $J^{\mu}$ are given by
\begin{equation}
A_{\mu}=A_{0}(x^{i})\delta^{0}_{\mu},
\label{Astatic}
\end{equation}
\begin{equation}
J^{\mu}=\frac{\sigma (x^{i})}{V}\delta^{\mu}_{0},
\label{Jstatic}
\end{equation}
with $i=1,2,3$. The corresponding expression for $u^{\mu}$ is
\begin{equation}
u^{\mu}=\frac{1}{V}\delta^{\mu}_{0},
\label{ustat}
\end{equation}
which clearly satisfies the constraint $u^{\mu}u_{\mu}=-1$.

Using the definition of the Maxwell tensor
\begin{equation}
F_{\mu\nu}=\partial_{\mu}A_{\nu}-\partial_{\nu}A_{\mu}
\label{maxdef}
\end{equation}
and Eqs. (\ref{confor})-(\ref{Jstatic}),
we see that Eq. (\ref{max}) contains only one non-trivial
equation:
\begin{equation}
\frac{1}{\sqrt{h}}\;\partial_{j}\left(\sqrt{h}\; h^{jk}\;
\frac{\partial_{k}A_{0}}
{V^{2}}\right)=\frac{4\pi \sigma}{V^3},
\label{nontriv}
\end{equation}
where $h$ and $h^{ij}$ are the determinant and the inverse of $h_{ij}$,
respectively.

Using Eq. (\ref{confor}) we determine the components of the Einstein tensor:
\begin{equation}
G_{00}=-3V^{2}h^{ij}\partial_{i}V\partial_{j}V+2V^{3}\nabla^{2}_{(h)}V+
\frac{1}{2}V^{4}R_{(h)},
\label{G00}
\end{equation}
\begin{equation}
G_{0i}=0,
\label{G0i}
\end{equation}
\begin{equation}
G_{ij}=R_{(h)ij}-\frac{2}{V^{2}}\partial_{i}V\partial_{j}V+
h_{ij}\left(\frac{1}{V^{2}}h^{mk}\partial_{m}V\partial_{k}V-\frac{1}{2}R_{(h)}\right),
\label{Gij}
\end{equation}
where $\nabla^{2}_{(h)}$, $R_{(h)}$ and $R_{(h)ij}$ denote the Laplacian 
operator,
the Ricci scalar and the Ricci tensor associated to the background space
with metric $h_{ij}$.

A simple procedure yields the components $E_{\mu\nu}$ and $M_{\mu\nu}$
in this static case.
The resulting expressions are given by
\begin{equation}
E_{00}=\frac{1}{8\pi}V^{2}h^{ij}\partial_{i}A_{0} \partial_{j}A_{0},
\label{E00}
\end{equation}
\begin{equation}
E_{0i}=0,
\label{E0i}
\end{equation}
\begin{equation}
E_{ij}=\frac{1}{4\pi}\left(-\frac{1}{V^{2}}\partial_{i}A_{0}\partial_{j}A_{0}+
\frac{1}{2}\frac{1}{V^{2}}h_{ij}h^{kl}\partial_{k}A_{0} 
\partial_{l}A_{0}\right),
\label{Eij}
\end{equation}
and
\begin{eqnarray}
M_{00}=\rho V^{2},&
M_{0i}=0,&
M_{ij}=0.
\label{Mcomps}
\end{eqnarray}

The above results allow the explicit determination of
Eqs.  (\ref{eins}).
Assuming that the background space is Euclidean,
we obtain the seven non-trivial equations
\begin{equation}
-3V^{2}h^{ij}\partial_{i}V\partial_{j}V+2V^{3}\nabla^{2}_{(h)}V=
V^{2}h^{ij}\partial_{i}A_{0}\partial_{j}A_{0}+8\pi\rho V^{2},
\label{eins00}
\end{equation}
\begin{equation}
-2\partial_{i}V\partial_{j}V+h_{ij}h^{kl}\partial_{k}V\partial_{l}V=
-2\partial_{i}A_{0}\partial_{j}A_{0}+h_{ij}h^{kl}\partial_{k}A_{0}\partial_{l}A_{0}.
\label{einsij}
\end{equation}
Contracting the last equation with $h^{ij}$ and using $h^{i}_{i}=3$ we 
find that
\begin{equation}
h^{kl}\partial_{k}V\partial_{l}V=h^{kl}\partial_{k}A_{0}\partial_{l}A_{0}.
\label{cuads}
\end{equation}
Combining this result with Eq. (\ref{einsij}) we get the equation
\begin{equation}
\partial_{i}V\partial_{j}V=\partial_{i}A_{0}\partial_{j}A_{0}
\label{ijred}
\end{equation}
which can be integrated to obtain
\begin{eqnarray}
A_{0}=\kappa V,&
\kappa^{2}=1,
\label{choice}
\end{eqnarray}
provided we take the additive integration constant equal to zero.
This result allows a  great simplification of
Eq. (\ref{eins00}) as well. In fact, using Eq. (\ref{choice}) to eliminate
$A_{0}$ in Eq. (\ref{eins00}), and combining the resulting expression with
the identity
\begin{equation}
\nabla^{2}_{(h)}V=\frac{2}{V}h^{ij}\partial_{i}V\partial_{j}V
-V^{2}\nabla^{2}_{(h)}\left(\frac{1}{V}\right)
\label{identi}
\end{equation}
we get the Poisson-type equation
\begin{equation}
\nabla^{2}_{(h)}\lambda+4\pi\rho\lambda^{3}=0,
\label{nonlin}
\end{equation}
where $\lambda=\frac{1}{V}$.

We can use Eq. (\ref{choice}) to eliminate $A_{0}$ in
the non-trivial Maxwell equation (\ref{nontriv}) as well. The outcoming 
expression is
\begin{equation}
\nabla^{2}_{(h)}\lambda + \frac{4\pi\sigma}{\kappa}\lambda^{3}=0.
\label{nonlinmax}
\end{equation}
Comparing Eqs. (\ref{nonlin}) and (\ref{nonlinmax}) we conclude that 
$\sigma$ and $\rho$
are related by
\begin{equation}
\sigma=\kappa \rho.
\label{dasrel}
\end{equation}

Equation (\ref{dasrel}) characterises the MP class of static solutions
for the EMCD equations,
in which the electrostatic repulsion between charges with the same sign 
is exactly
balanced by the gravitational attraction within the fluid. This kind of
fluid has been named {\it electrically counterpoised dust} \cite{bw1}. 
Such a precise
equilibrium of gravity and electricity is physically possible and can be 
carried out
with slightly ionised Hidrogen \cite{bw2,bo2}.

We remark that the above analysis of the EMCD equations is a generalisation
of the one presented by  Lynden-Bell {\it et al.} \cite{lyn1},
who made the Cartesian choice $h_{ij}=\delta_{ij}$ for the background 
space metric.

If we
assume $\rho=0$, then
Eq. (\ref{nonlin})
reduces to the usual  Laplace equation $\nabla^{2}_{(h)}\lambda=0$ and the
electrovac, multi-black hole solution
follows straightforwardly. Assuming spherical symmetry
and using spherical coordinates $(x^0=t, x^1=r, x^2=\theta, x^3=\phi)$, 
we find
\begin{equation}
\lambda=1+\frac{m}{r}.
\label{esfesol}
\end{equation}
In the far-asymptotic region, the behaviour of this solution is 
approximately
given by
\begin{eqnarray}
V \approx 1-\frac{m}{r}, &
g_{00} \approx -1+\frac{2m}{r}, &
A_{0} \approx \pm (1-\frac{m}{r}).
\label{assymp1}
\end{eqnarray}
The corresponding expression for the electric field is
\begin{equation}
F_{01} \approx \frac{q}{r^2},
\label{assymp2}
\end{equation}
where
\begin{equation}
q=\mp m.
\label{qigualm}
\end{equation}

Equation (\ref{confor})
implies that
the invariant area of any 2-sphere surrounding the origin
is given by
$\frac{4\pi r^{2}}{V(r)^{2}}$. Therefore,
the set $r=0$, $t=constant$
has a non-zero invariant area given by $4\pi m^{2}$.
In fact, a simple coordinate transform shows that the null hypersurface 
$r=0$
is the horizon of the extremal Reissner-Nordstr\"{o}m solution.
Also, if we define the new radial coordinate $\tilde{r}=-r$
and perform the standard analysis \cite{haha},
then we find that this horizon encloses a point-like, essential singularity
placed at  $\tilde{r}=m$. In fact, the invariant area vanishes and the
scalar
$F_{\mu\nu}F^{\mu\nu}=\lambda^{-4}\left(\frac{d\lambda}{dr}\right)^{2}$
blows up at that point.

\section{The Linear Model}

Equations (\ref{nonlin}) and (\ref{dasrel}) were
discussed by Das \cite{das} in his study of equilibrium configurations of
self-gravitating charged dust.
More recently, G\"{u}rses \cite{gur} has considered
non-electrovac solutions when Eq. (\ref{nonlin}) is linear and homogeneous.
This situation corresponds to his choice 
$\rho=\frac{b^{2}}{4\pi\lambda^{2}}$
for constant $b$.
In this case, Eq. (\ref{nonlin}) admits the solution
\begin{equation}
\lambda=\frac{a\sin \left(br\right)}{r},
\label{gursol}
\end{equation}
where $a$ is an integration constant.
(The signs of $a$ and $b$ are not relevant to the geometry
and we assume $a>0$, $b>0$.)

The oscillatory behaviour of this solution
implies a geometry with complicated radial dependence. In fact, the
invariant area vanishes for an infinite set of values of $r$, and the
Ricci scalar $R=\frac{2 b^2 r^2}{a^2 \sin\left(br\right)^2}$ blows up 
wherever the
invariant area vanishes, except for $r=0$.

Hartle and Hawking \cite{haha}
showed that the multi-black hole geometry (\ref{mpsol}) is the only 
electrovac
solution without nacked singularities. This result motivated G\"{u}rses 
\cite{gur}
to consider the match of arbitrary electrovac solutions to charged dust 
solutions,
so that the horizons and curvature singularities of the external 
solutions can be
eliminated. Nevertheless, the internal solution must be chosen with care 
so that new
singularities are not included in the geometry. The oscillatory, point-like
curvature singularities of the linear model (\ref{gursol})
constitute a clear-cut example of this potential trouble.

We develope a new set of parameters to analyse the geometry proposed by 
G\"{u}rses.
These parameters allow simple choices for the charged dust solution, so 
that internal
curvature singularities can be avoided.

In order to construct a spherically symmetric,
class $C^1$ model we require the continuity of the metric and its first
radial derivative at the boundary of the dust distribuition. In other words,
we want $r\lambda$ and $\left(r\lambda\right)^\prime$
to be continuous at the coordinate radius $r_1>0$ of the charged source.
Imposing these matching conditions to
the internal and external solutions (\ref{gursol}) and (\ref{esfesol}),
we obtain
\begin{equation}
\tan \beta = \left(1+\mu\right) \beta,
\label{tan}
\end{equation}
\begin{equation}
\alpha=\sqrt{\left(1+\mu\right)^2+\frac{1}{\beta^2}},
\label{alp}
\end{equation}
where
\begin{eqnarray}
\beta=b r_1,&
\mu=\frac{m}{r_1},&
\alpha=\frac{a}{r_1}.
\label{para1}
\end{eqnarray}
In terms of the new adimensional parameters, the internal and external 
solutions are
given by
\begin{eqnarray}
\lambda_{I}=\frac{\alpha \sin \left(\beta x\right)}{x},&
0 \le x \le 1,
\label{soladim1}
\end{eqnarray}
\begin{eqnarray}
\lambda_{II}=1+\frac{\mu}{x},&
1 \le x < \infty,
\label{soladim2}
\end{eqnarray}
where $x=\frac{r}{r_1}$. Given a particular value of $\mu>0$, each of the
infinite positive roots of Eq. (\ref{tan}) defines a family of 
asymptotically flat
solutions. These solutions are singularity-free only if $\beta<\pi$.
Clearly, only one of the positive roots of Eq. (\ref{tan}) satisfies this
condition, so that every acceptable family of solutions is defined by 
some $\beta$
in the interval $\left(0,\frac{\pi}{2}\right)$.

As in the spherically symmetric case considered by Bonnor and 
Wickramasuriya \cite{bw2},
the linear model implies arbitrarily large redshifts when the mass
parameter $\mu$ tends to infinity.
Let us consider two spatial points $P$ and $Q$ at rest in the 
coordinates of the
conformastatic metric (\ref{confor}).
If we take $Q$ at spatial infinity, the redshift of light emitted at $P$ 
and received at $Q$ is
$\lambda(P)-1$.
Using Eq. (\ref{soladim1}), we find that $\lambda_{I}(r=0)=\alpha \beta$,
so that the maximum redshift attainable with this
bounded source is
\begin{equation}
Z=\alpha \beta - 1.
\label{redshift}
\end{equation}
Using Eqs. (\ref{tan}) and (\ref{alp}), we find that $\beta \approx 
\frac{\pi}{2}$,
$\alpha \approx \mu$, and $Z$ is unbounded when $\mu$ is arbitrarily large.

\section{The -Sine-Gordon Model}
The non-linear potential equation (\ref{nonlin}) takes the spherically
symmetric form
\begin{equation}
\frac{d^{2}\lambda}{dr^{2}}+\frac{2}{r}\frac{d\lambda}{dr}+4\pi\rho\lambda^{3}=0.
\label{equesfe}
\end{equation}
Using the new radial coordinate $\tau=\frac{1}{r}$, the same differential
equation can be written as
\begin{equation}
\frac{d^{2}\lambda}{d\tau^{2}}+\frac{4\pi\rho}{\tau^{4}}\lambda^{3}=0.
\label{equesfetau}
\end{equation}
If $\rho$ and $\lambda$ satisfy the condition
\begin{equation}
\rho=\frac{\delta^2}{4\pi}\frac{\tau^{4}\sin\lambda}{\lambda^{3}},
\label{newfun}
\end{equation}
then (\ref{equesfetau}) finally reduces to the
-sine-Gordon equation
\begin{equation}
\frac{d^{2}\lambda}{d\tau^{2}}+\delta^2 \sin\lambda=0.
\label{soliton0}
\end{equation}
It admits the solution
\begin{equation}
\lambda\left(\tau\right) =
2\arcsin\left[\tanh\left(\delta\tau+c\right)\right],
\label{newsol1}
\end{equation}
where $c$ is an integration constant,
and $\delta$ is assumed to be positive.
In terms of the original radial coordinate, this solution read
\begin{equation}
V\left(r\right)=
\frac{1}{2\arcsin\left[\tanh\left(\frac{\delta}{r}+c\right)\right]}.
\label{newsol2}
\end{equation}
We observe that $V(0)^{2}$ is finite. Hence the
invariant area vanishes for $r=0$, and the set $r=0$, $t=constant$ is
point-like with respect to this solution.
A numerical study of the invariants $F_{\mu\nu}F^{\mu\nu}$, $R$,
$R^{\alpha\beta}R_{\alpha\beta}$,
$R^{\alpha\beta\gamma\delta}R_{\alpha\beta\gamma\delta}$,
$R_{\alpha\beta\gamma\delta}R^{\gamma\delta}{}_{\sigma\rho}R^{\alpha\beta\sigma\rho}$
indicates that these
quantities are bounded for non-negative $r$, whenever $c$ is positive.

If we choose
\begin{equation}
c=\frac{1}{2}\ln\left[\frac{1+\sin(1/2)}{1-\sin(1/2)}\right],
\label{cchoice}
\end{equation}
then the far-asymptotic $(r \rightarrow \infty)$
behaviour of this solution is given by
Eqs. (\ref{assymp1})-(\ref{qigualm})
with $m=2\delta\cos(1/2)$. Therefore, this charged dust solution
is asymptotically flat, exactly as the previously considered electrovac 
solution.
Also, Eqs. (\ref{newsol1}), (\ref{cchoice}), and (\ref{newfun}) imply 
that the
dust energy density of this model is definite positive for $r>0$, 
provided $m>0$.

The coordinate transform $\tilde{r}=-r$ reveals the existence of a
point-like, essential singularity at $\tilde{r}=\frac{\delta}{c}$. In 
fact, the
invariant $F_{\mu\nu}F^{\mu\nu}$ blows up at this point. Additionally, 
this geometry is
asymptotically flat (with negative mass) for $\tilde{r} \rightarrow \infty$.

The above considerations suggest a division of
the spacetime manifold into three parts, separated by the point-like
singularity placed at $\tilde{r}=\frac{\delta}{c}$ and the time-like
world-line $\left(r=0, t\right)$.
We have seen that two of these parts are asymptotically flat. The third one
is defined by $\tilde{r} \in \left(0,\frac{\delta}{c}\right)$ and has
spatial volumen $\Omega$ given by the finite integral
\begin{equation}
\Omega= 4 \pi \int\limits_{0}^{\frac{\delta}{c}} 
\frac{\tilde{r}^2}{V(\tilde{r})^3} d\tilde{r}.
\label{vol}
\end{equation}

Although the structure of this charged dust solution is complicated by the
juxtaposition of bounded and unbounded spatial regions, we see that the 
far-asymptotic
behaviour is its most relevant feature. In fact, a singularity-free
class $C^1$ geometry can be built if we match the metric (\ref{newsol1})
to the G\"{u}rses internal solution at $r=r_1>0$.
The result will be a composite sphere of charged dust which extends to 
infinity.
(In this and the following compound source models, the sign of $\sigma$ 
is assumed to be
the same everywhere.)

The adimensional form of the new solution is given by
\begin{eqnarray}
\lambda_{I}= \frac{\alpha \sin \left( \beta x \right) }{x} , &
0 \le x \le 1,
\label{lambdaI2}
\end{eqnarray}
\begin{eqnarray}
\lambda_{II}= 2 \arcsin \left[ \tanh \left( \frac{\gamma \mu}{x} + c 
\right) \right] , &
1 \le x < \infty,
\label{lambdaII2}
\end{eqnarray}
where $\gamma=\frac{1}{2 \cos \left(1/2 \right)}$. In this case, the 
$C^1$ match of $\lambda_{I}$
and $\lambda_{II}$ implies the equations
\begin{equation}
\tan \beta  = f \left(\mu \right) \beta,
\label{tan2}
\end{equation}
\begin{equation}
f \left(\mu \right)= \frac{1}{ 1- \frac{\gamma \mu  \sqrt{1-\chi^2 } } 
{\arcsin \chi} },
\label{fdef}
\end{equation}
and
\begin{equation}
\alpha = 2 \sqrt{
\arcsin \left(\chi\right)^{2} +\frac{1}{\beta^2}\left[ \arcsin \chi
-\gamma\mu\sqrt{1-\chi^{2}}\right]^{2}
},
\label{alfa2}
\end{equation}
where $\chi=\tanh \left( \gamma \mu +c \right)$.

As in the linear model, each value of the mass parameter $\mu>0$ 
determines an infinite set of
positive roots for equation (\ref{tan2}). Once again, only the root
satisfying $0<\beta<\frac{\pi}{2}$ defines a family of singularity-free, 
asymptotically flat solutions.
  This family is completely fixed when we evaluate $\alpha$ with Eq. 
(\ref{alfa2}).

The determination of $\beta$ and $\alpha$ follows essentially the same
procedure in the linear and -sine-Gordon models.
Nevertheless, the implicitly defined function $\beta\left(\mu\right)$ is 
remarkably different in these
two cases. We see that the coefficient of $\beta$ in the right-hand side 
of Eq. (\ref{tan}) is a linear function
of $\mu$. On the other hand, Eq. (\ref{fdef}) reveals a totally 
different structure in $f \left(\mu \right)$,
which turns out to be bounded for $\mu$ in the interval 
$\left[0,\infty\right)$.

The greatest redshift obtainable from a -sine-Gordon solution can be 
easily found in the limit
of arbitrarily large $\mu$. The definition of $\chi$ and Eq. 
(\ref{alfa2}) allow the approximations
$\chi\approx 1$, $\sqrt{1-\chi^2} \approx 2 
e^{-\left(\gamma\mu+c\right)}$, and
$\alpha \beta \approx \pi \sqrt{1+\beta^2}$
for large $\mu$. Equations (\ref{tan2}) and (\ref{fdef}) imply that 
$\beta$ tends to zero,
and the maximum redshift falls to
\begin{equation}
Z = \pi-1 = 2.1415....
\label{maxZ2}
\end{equation}
in this limit.

\section{Sources with Non-Noetherian Symmetries}
We have seen that G\"{u}rses assumed the equation 
$\rho=\frac{b^2}{4\pi\lambda^2}$ to
  discuss the linear, homogeneous  case of the Poisson-type equation 
(\ref{nonlin}).
Analogously, Varela \cite{var}
considered other choices for $\rho$, leading to non-linear Poisson-type 
equations which are well known
in Soliton Physics. The choice $\rho=\frac{\delta^2 
\tau^4}{4\pi}\frac{\sin\lambda}{\lambda^3}$ is the
starting point for the -sine-Gordon model developed in the previous 
Section. This {\it ad-hoc} functional
relationship between $\rho$ and the metric function $\lambda$ is 
justified only by the interest of the
analytic integral of Eq. (\ref{nonlin}), given by Eq. (\ref{newsol1}), 
and the associated spacetime geometry.

A completely different method for constructing sources for 
Majumdar-Papapetrou spacetimes comes out when
we assume the symmetry
\begin{equation}
\pounds_{\xi} \, g_{\mu\nu}=\psi g_{\mu\nu},
\label{lieder}
\end{equation}
where the left-hand side is the Lie derivative of the metric along
the congruence generated by the vector field $\xi$, and
$\psi$ is an arbitrary function of $x^{\alpha}$.
Using the conformastatic metric (\ref{metric}) with $V=V(r)$,
and choosing
\begin{equation}
\xi=\xi^{0} \frac{\partial}{\partial t} + \xi^{1} 
\frac{\partial}{\partial r} ,
\label{xiform}
\end{equation}
with $\xi^{0}=\xi^{0}(r)$ and $\xi^{1}=\xi^{1}(r)$,
we get the following system of equations:
\begin{equation}
\frac{\partial V}{\partial r} \xi^{1} = \frac{\psi}{2} V,
\label{ld00}
\end{equation}
\begin{equation}
V^2 \frac{\partial \xi^{0}}{\partial r} = 0,
\label{ld01}
\end{equation}
\begin{equation}
-\frac{\partial V}{\partial r} \xi^{1} + \frac{\partial 
\xi^{1}}{\partial r} V = \frac{\psi}{2} V,
\label{ld11}
\end{equation}
\begin{equation}
\left(-r\frac{\partial V}{\partial r}  + V \right) \xi^{1} = r 
\frac{\psi}{2} V.
\label{ld22}
\end{equation}
Solving Eqs. (\ref{ld00})-(\ref{ld22}), we find that $\xi^{0}$ and 
$\psi$ are constants, and
\begin{equation}
\xi^{1}=\psi r,
\label{ldsol1}
\end{equation}
\begin{equation}
V=D \sqrt{r},
\label{ldsol2}
\end{equation}
where $D$ is another integration constant.
This homothetic vector field is naturally extended to
\begin{equation}
\xi=\upsilon+\psi r \frac{\partial}{\partial r},
\label{homogen}
\end{equation}
where $\upsilon$ is any linear combination -with constant coefficients- 
of $\frac{\partial}{\partial t}$
and the three generators of spatial rotations.

The positive definite
energy density of this homothetic charged dust sphere  can be easily 
obtained if we combine
Eqs. (\ref{nonlin}) and (\ref{ldsol2}):
\begin{equation}
\rho= \frac{D^2}{16 \pi r}=\frac{D^4 \lambda^2}{16 \pi}.
\label{ldrho}
\end{equation}
The charge density is derived from Eq. (\ref{dasrel}).

This family of solutions describes homothetic charged dust with a point-like
curvature singularity at $r=0$. It was originally discussed by Herrera 
and Ponce de Le\'{o}n \cite{HePon},
who used the most general static, spherically symmetric line element 
expressed in Schwarzschild coordinates.
In fact, these authors obtained it as a member of a class of solutions 
of the Einstein-Maxwell
equations admitting conformal motions. What we find out is that the less 
general conformastatic line element
(\ref{metric}) fits the geometrical symmetry (\ref{lieder}) uniquely, 
i.e., the combination of these two
equations singles out a particular family of sources with metric 
(\ref{ldsol2}),
and definite positive density (\ref{ldrho}).

We can use this homothetic dust solution to construct an asymptotically 
flat, singularity-free,
MP spacetime. In order to eliminate the singularity at $r=0$,
we match the homothetic and G\"urses geometries at $r=r_{1}>0$. Then the 
homothetic and electrovac solutions
are matched across the surface $r=r_{2}$, with $r_2>r_1$.
As in the -sine-Gordon model, this source is made up of
two different types of charged dust. We point out that Herrera and Ponce 
de Le\'{o}n matched
the homothetic geometry to the interior Schwarzschild solution. This 
different choice led them to a source
with a core composed of a neutral perfect fluid with non-vanishing 
pressure.

The $C^1$ match of metrics (\ref{gursol}) and (\ref{ldsol2}) at $r=r_1$, 
and metrics (\ref{ldsol2}) and
(\ref{esfesol}) at $r=r_2$ provides us with four conditions, i.e.,
\begin{equation}
D=\frac{1}{2\sqrt{r_2}},
\label{eqD}
\end{equation}
\begin{equation}
m=r_2,
\label{eqm}
\end{equation}
\begin{equation}
a\sin\left(b r_1\right)=2\sqrt{r_1 r_2},
\label{eqa}
\end{equation}
\begin{equation}
a b \cos\left(b r_1\right)=\sqrt{\frac{r_2}{r_1}}.
\label{eqb}
\end{equation}
We observe that the total mass of this solution equals the radius of the
composite dust sphere in these coordinates.
However, the ratio of $m$ to $r_1$ can be used to label the arising 
families of geometries.
This fact stablishes an interesting link with the linear and 
-sine-Gordon models,
in which $\mu=\frac{m}{r_1}$
is the fundamental parameter.
Using the parameters $\alpha$ and $\beta$ defined in Eq. (\ref{para1}),
and the independent variable $y=\frac{r}{r_2}$,
we obtain the adimensional form of this solution:
\begin{eqnarray}
\lambda_I=\frac{\alpha \sin\left(\beta \mu y \right)}{\mu y},&
0 \le y \le \frac{1}{\mu},
\label{newgursol}
\end{eqnarray}
\begin{eqnarray}
\lambda_{II}=\frac{2}{\sqrt{y}},&
\frac{1}{\mu} \le y \le 1,
\label{newhomosol}
\end{eqnarray}
\begin{eqnarray}
\lambda_{III}=1+\frac{1}{y},&
1 \le y < \infty.
\label{newevac}
\end{eqnarray}
We choose $\beta$ as the root of $\tan\beta=2\beta$ in the interval
$\left(0,\frac{\pi}{2}\right)$, and note that $\beta$ does not depend on 
$\mu$
in this case.
The remaining parameter $\alpha$ is calculated with the expression
\begin{equation}
\alpha=\sqrt{\mu \left(4+\frac{1}{\beta^2}\right)}.
\label{eqalfa}
\end{equation}

The simplicity of this method for constructing homothetic MP spacetimes 
encourages us to
study other non-Noetherian symmetries. Our attention focuses on the 
coupling of charged
matter with geometry, and we expect the higher symmetries of the Ricci and
  energy-momentum tensors to play an important role in the conformation 
of the sources.

Ricci collineations (RCs) \cite{rc},
\begin{equation}
\pounds_{\eta}\,R_{\beta\gamma}=0,
\label{rc}
\end{equation}
are interesting as the Ricci tensor is the trace
of the curvature tensor, which is derived from
the connection. Therefore the study of RCs
has a natural geometrical significance \cite{hall}.

On the other hand, the energy-momentum tensor represents the
external sources of gravity
and its symmetries seem more relevant from the physical point of view.
Thus the study of matter collineations (MCs),
\begin{equation}
\pounds_{\zeta}\,T_{\beta\gamma}=0,
\label{mc}
\end{equation}
should be pertinent to the construction of charged sources.

Studies of this symmetry have been carried out in various contexts.
Sharif \cite{sha} has looked at MCs in Bianchi and Kantowski-Sachs 
spacetimes,
Carot {\it et al.} \cite{carot} have considered its algebraic properties
in the cases of degenerate and non-degenerate energy-momentum tensor,
and Hall {\it et al.} \cite{hall} have treated
the similarities between MCs and RCs within a more general framework.

We point out that MCs  can be derived from 
$\pounds_{\zeta}\,G_{\beta\gamma}=0$
instead of Eq. (\ref{mc}). Obviously, the equivalence of these two 
equations is
guaranteed by Eq. (\ref{eins}).

The analysis of MCs is surprisingly simple when we assume Eq. (\ref{metric})
with $V=V(r)$. The homothetic vector (\ref{homogen}) turns out to be
a MC when $V=D \sqrt{r}$. Due to its relationship with a lower level 
symmetry,
this MC is said to be {\it improper}. Other solutions of Eq. (\ref{mc}) can
be found when $\zeta$ takes the form
\begin{equation}
\zeta=\zeta^{0}(t,r) \frac{\partial}{\partial t} + \zeta^{1}(t,r) 
\frac{\partial}{\partial r}.
\label{zeta}
\end{equation}
In this case, Eq. (\ref{mc}) entails the following system of
partial differential equations:
\begin{equation}
\frac{\partial Q}{\partial r} \zeta^1 + 2 Q \frac{\partial 
\zeta^0}{\partial t}=0,
\label{mc11}
\end{equation}
\begin{equation}
P \frac{\partial \zeta^1}{\partial t}+Q \frac{\partial \zeta^0}{\partial 
r}=0,
\label{mc12}
\end{equation}
\begin{equation}
\frac{\partial P}{\partial r} \zeta^1 + 2 P \frac{\partial 
\zeta^1}{\partial r}=0,
\label{mc22}
\end{equation}
\begin{equation}
\left(2 P + r \frac{\partial P}{\partial r} \right) \zeta^1 = 0,
\label{mc33}
\end{equation}
where
\begin{equation}
P=\frac{1}{V^2}\left(\frac{\partial V}{\partial r} \right)^2,
\label{defP}
\end{equation}
\begin{equation}
Q=\frac{V^2}{r}\left[3r\left(\frac{\partial V}{\partial r} \right)^2 - 
2r V \frac{\partial^2 V}{\partial r^2}
- 4V \frac{\partial V}{\partial r} \right].
\label{defQ}
\end{equation}

Assuming $\zeta^1 \ne 0$, Eqs. (\ref{mc33}) and (\ref{defP}) can be 
solved to obtain
\begin{equation}
V=H r^\epsilon,
\label{newv}
\end{equation}
where $H$ and $\epsilon$ are integration constants.
Combined with Eq. (\ref{nonlin}), this result yields the charged dust 
energy density
\begin{equation}
\rho=\frac{\epsilon\left(1-\epsilon\right) H^2}{4\pi 
r^{2\left(1-\epsilon\right)}}
=\frac{\epsilon\left(1-\epsilon\right) H^{\frac{2}{\epsilon}} 
\lambda^{2\left(\frac{1}{\epsilon}-1\right)}}{4\pi},
\label{newrho}
\end{equation}
which is definite positive only if $\epsilon \in \left(0,1\right)$.
Using Eqs. (\ref{defP})-(\ref{newv}), we calculate $P$ and $Q$ and proceed
to solve
Eq. (\ref{mc22}) for $\zeta^1$ and then
Eqs. (\ref{mc11}) and (\ref{mc12}) for $\zeta^0$. The straightforward 
evaluation of the arising
integration functions provides us with the results
\begin{equation}
\zeta=\frac{\ln r}{3 H^4}\frac{\partial}{\partial t} + rt 
\frac{\partial}{\partial r},
\label{ep12}
\end{equation}
for $\epsilon=\frac{1}{2}$, and
\begin{equation}
\zeta=\left[\frac{\epsilon}{ H^4 \left(\epsilon-2\right) 
\left(4\epsilon-2\right) r^{ \left(4\epsilon-2\right) } }
+ \left( \frac{1}{2}-\epsilon \right) t^2 \right] 
\frac{\partial}{\partial t} + rt \frac{\partial}{\partial r}
\label{epne12}
\end{equation}
otherwise.

The Kretschmann scalar derived from (\ref{newv}) is given by
\begin{equation}
R^{\alpha\beta\gamma\delta}R_{\alpha\beta\gamma\delta}=
\frac{4 H^4 \epsilon^2 \left(7\epsilon^2-12\epsilon+7\right)}{r^{4 
(1-\epsilon)}} ,
\label{kret}
\end{equation}
and the invariant area of any 2-sphere which is inmerse in the fluid
and sorrounds the origin of coordinates
is $\frac{4 \pi r^{2(1-\epsilon)}}{H^2}$. Thus the point-like curvature 
singularity
at $r=0$ exists for every $\epsilon \in \left(0,1\right)$.

We match our MC solution (\ref{newv}) to the G\"{u}rses and electrovac 
geometries
at $r=r_1$ and $r=r_2>r_1$, respectively. The arising class $C^1$, 
asymptotically flat solution
is given by
\begin{eqnarray}
\lambda_I=\frac{\alpha \sin\left[\left(\frac{1}{\epsilon}-1\right)\beta 
\mu y \right]}
{\left(\frac{1}{\epsilon}-1\right)\mu y},&
0 \le y \le \frac{1}{\left(\frac{1}{\epsilon}-1\right)\mu},
\label{collsol1}
\end{eqnarray}
\begin{eqnarray}
\lambda_{II}=\frac{1}{\left(1-\epsilon\right) y^\epsilon},&
\frac{1}{\left(\frac{1}{\epsilon}-1\right)\mu} \le y \le 1,
\label{collsol2}
\end{eqnarray}
\begin{eqnarray}
\lambda_{III}=1+\frac{\epsilon}{\left(1-\epsilon\right) y},&
1 \le y < \infty,
\label{collsol3}
\end{eqnarray}
where
\begin{equation}
\alpha=\left(\frac{1}{\epsilon}-1\right)^\epsilon \mu^\epsilon
\sqrt{\frac{1}{\left(1-\epsilon\right)^2} + \frac{1}{\beta^2}}.
\label{alfaeta}
\end{equation}
The total mass and the radius of the composite dust sphere are related by
\begin{equation}
\frac{m}{r_2}=\frac{\epsilon}{1-\epsilon}.
\label{mr2}
\end{equation}
This class of solutions is singularity-free whenever
$\beta$ is chosen as the root of
\begin{equation}
\tan\beta=\frac{\beta}{1-\epsilon}
\label{betaepa}
\end{equation}
within the interval $\left(0,\frac{\pi}{2}\right)$
for $\epsilon \in \left(0,1\right)$.

The study of RCs in MP spacetimes has not been carried out very far;
however, some points have been noted.
A similar method can be used to solve  Eq. (\ref{rc}) and construct 
charged dust sources
admitting RCs with $(t,r)$ dependence. The system of equations 
corresponding to
Eqs. (\ref{mc11})-(\ref{mc33}) has a more complicated dependence on $V$
and its derivatives, and a third order differential equation must be solved
to find the metric function
\begin{equation}
V=K r^\varepsilon e^\frac{\omega}{r},
\label{vrc}
\end{equation}
where $K$, $\varepsilon$, and $\omega$ are integration constants.
The evaluation of integration functions forces us to choose $\omega=0$,
so that Eq. (\ref{vrc}) actually reduces to the MC metric, given by Eq. 
(\ref{newv}).

\section{Final Remarks}
We have constructed several types of sources for the Einstein-Maxwell field
in the MP case of the theory. These are static charged dust spheres for 
which
the energy and charge densities satisfy Eq. (\ref{dasrel}). In all 
cases, the central
region is described by the G\"{u}rses solution which is non-singular 
whenever the
values of the parameters are appropiately fixed.

The linear and -sine-Gordon models are asymptotically flat and have the 
same $r^{-2}$
dependence in the electric field at large distances from the symmetry 
center. The linear model
assumes a bounded charge distribution, and the -sine-Gordon geometry 
describes charged dust which
extends to infinity. This unbounded source becomes dilute at infinity 
fast enough
so that the total mass (and charge) is finite \cite{beig}.

Although both geometries have the same asymptotic properties, an 
observer at spatial
infinity can discrimininate between them. To this end, the observer must 
measure the maximum
redshift attainable from these sources for very large values of $\mu$.

The study of homothetic charged dust sources for MP spacetimes led us to 
a solution previously
found by Herrera and Ponce de Le\'{o}n in Schwarzschild coordinates.
The simplicity of our calculation
in isotropic coordinates pointed out an interesting affinity
between this symmetry and the conformastatic metric with $V=V(r)$.
This observation motivated the consideration of higher
non-Noetherian symmetries in MP spacetimes.

We have solved the MC equations (\ref{mc}) to find a class of
MP solutions with positive dust energy density,
which includes the homothetic solution as a particular case.
The analytical forms of the associated vector fields indicate that the 
homothetic solution
plays a distinctive role within this class. Preliminary calculations 
suggest that
these MCs are {\it proper} (not derived from any lower symmetry), but
additional work is due in order to clarify this aspect of the solutions.

The above observations encourage us to investigate the whole structure 
of the MCs
in spherically symmetric MP spacetimes. The fact that the
energy-momentum tensor is non-degenerate will determine a finite 
dimensional Lie algebra
of MCs \cite{sha,carot} associated to the metric (\ref{newv}).
A similar study of the corresponding RCs could be carried out in detail
and might prove interesting.
This future work should convey us to a classification scheme for MP 
solutions based on
non-Noetherian symmetries.

\section*{Acknowledgments}

The author acknowledges the use of the MAPLE program and the TENSOR package
at various stages of this work.

\end{document}